\documentclass[aps,preprint,showpacs,preprintnumbers,amsmath,amssymb]{revtex4}

\newcommand{\be}{\begin{equation}}
\newcommand{\ee}{\end{equation}}

\begin{document}

%\input{epsf}
%\draft

\title{Effects of anisotropy on a quantum Heisenberg spin glass 
in a three-dimensional hierarchical lattice}

\author{ \sc J. Ricardo de Sousa }
\affiliation{Departamento de F\'{\i}sica
 Universidade do Amazonas\\
3000 Japiim,
69077-000, Manaus, AM  - Brazil}

\author{ \sc {Beatriz Boechat, Claudette Cordeiro}} 
\affiliation{Instituto de   F\'{\i}sica - Universidade Federal
Fluminense \\ 
Av. Litor\^anea s/n,  Niter\'oi, 24210-340, RJ, Brazil}

\author{ \sc N. S. Branco }
\affiliation{Departamento de F\'{\i}sica,
Universidade Federal de Santa Catarina, 
88040-900, Florian\'opolis, SC  - Brazil \\
\email{nsbranco@fisica.ufsc.br}}

\date{\today}

\begin{abstract}

	We study the anisotropic Heisenberg spin-glass model in 
a three-dimensional hierarchical lattice (designed to approximate the
cubic lattice), within a real-space renormalization-group approach. 
Two different initial probability distributions for the exchange
interaction ($J_{ij}$), Gaussian and uniform, are used,
with zero mean and width $\bar{J}$.
The $(kT/\bar{J}) \times \Delta_0$ phase diagram is obtained, where
$T$ is the temperature, $\Delta_0$ is the first moment of the
probability distribution for the uniaxial anisotropy,
and $k$ is the Boltzmann constant. For the Ising
model ($\Delta_0=1$), there is a spin-glass phase at low
temperatures (high $\bar{J}$) and a paramagnetic phase at high
temperatures (low $\bar{J}$).
For the isotropic Heisenberg model ($\Delta_0=0$), our results indicate
no spin-glass phase at finite temperatures. 
The transition temperature between the
spin-glass and paramagnetic phase decreases with $\Delta_0$,
as expected, but goes to zero at a finite value of the
anisotropy parameter, namely $\Delta_0 = \Delta_c \sim 0.59$. Our results
indicate that the whole transition line, between the paramagnetic
and the spin-glass phases, for $\Delta_c \leq \Delta_0 < 1$,
belongs to the same universality class as the transition for
the Ising spin glass.

\noindent     

\end{abstract}

\maketitle

\newpage

\section{Introduction}

	Considerable attention during the last decade has been devoted to the
investigation of systems displaying spin-glass (SG) order. Numerical studies
\cite{numstud} have revealed that the SG phase transition occurs in the
three-dimensional Ising model (strongly anisotropic system),
indicating that the lower critical dimension for the Ising SG would be $d_l=2$.
On the other hand, many real materials that show SG order are
Heisenberg-like rather than Ising-like, in the sense that the magnetic
anisotropy is considerably weaker than the isotropic exchange interaction.
Some Monte Carlo simulations \cite{mc1,mc2} have indicated
that the isotropic three-dimensional classical Heisenberg SG with finite-range 
interaction does not
exhibit the conventional SG order at finite temperatures in zero field,
while Lee and Young \cite{leeyoung} found such an ordered phase at
finite temperatures for this model.

        Experiments clearly demonstrate the existence of order at finite
temperatures in
Heisenberg-like SG systems \cite{petit}, where the chirality-driven mechanism
proposed by Kawamura \cite{mc2,kawa} can be interpreted consistently to explain
some of the puzzles concerning the experimentally observed SG transition in
zero field \cite{petit}. Note that the numerical observation of a
finite-temperature chiral-glass (CG) transition ($T_{CG}>0$) in the 
three-dimensional classical
Heisenberg SG is not inconsistent with the earlier observation of the
absence of the conventional SG order at finite temperatures ($T_c=0$). 
In Refs. \onlinecite{mc2} and \onlinecite{kawa} it is
suggested that the SG-paramagnetic critical temperature obeys
$T_c<T_{CG}$, and quite 
possibly $T_c=0$ in three dimensions.
In the presence of a small random magnetic anisotropy, which always exists
in real experimental situations, an SG phase is expected to emerge as a
result of the fact that the anisotropy mixes the two degrees of freedom,
spin and chirality. Therefore, in the chirality-driven mechanism the SG
phase transition experimentally observed in a class of compounds such as
CuMn is essentially governed by the CG fixed point.

	Note, however, that some numerical results support that the SG 
transition temperature might coincide
with the CG transition temperature, i.e., $T_c=T_{CG}>0$\cite{leeyoung,naka}, 
in contrast with the results of Refs. \onlinecite{mc2} and 
\onlinecite{kawa} (which show that in three dimensions the spin and the
chirality are decoupled on sufficiently long length scales, with 
$T_c<T_{CG} $). Therefore, the presence of SG order in the three-dimensional
short-range Heisenberg SG is still an open question. In
four or more dimensions, there is numerical evidence of a phase transition
\cite{imaga}, and so the lower critical dimension $d_l$ for the short-range
Heisenberg spin glass should satisfy $3 \leq d_l<4$.

	On the other hand, various types of anisotropies have a profound 
influence on the SG phase such
as Dzyaloshinski-Moriya (DM), dipolar coupling, and uniaxial. 
A weak anisotropy is
crucially important in realizing a finite-temperature SG transition,
which causes a crossover from the isotropic Heisenberg behavior to
the anisotropic Ising behavior. The expected Heisenberg-to-Ising
crossover, however, has not been observed experimentally, and this puzzle
has remained unexplained. 
Using a hybrid Monte Carlo method in the
short-range $\pm J$ Heisenberg spin glass with random anisotropy 
of a DM type ($D$) on a
simple cubic lattice \cite{matsu1}, it has been shown that for small values of
$D$, the transition temperature vanishes as
$\frac{T_c}J \simeq 0.53\left( \frac{D}{J} \right)^{1/4}$. This result 
is consistent with those found by Morris \textit{et al.} \cite{morris} 
based on a scaling argument.
When the spin interactions are of long-range Ruderman-Kittel-Kasuya-Yoshida
(RKKY) type, the critical temperature has a much weaker dependence on the
anisotropy, namely $\frac{T_c}J \simeq $ $\left[ \ln (J/D)\right] ^{-1/2}$
($D<<<J$) \cite{bray}. These two studies indicate that in the isotropic 
limit ($D=0$) there is no SG order at finite temperatures, i.e, $T_c=0$.
However, this result has recently been challenged
\cite{leeyoung}. To the best of our knowledge, the only previous work
on a spin-glass model with uniaxial anisotropy is the one by Matsubara
\textit{et al.} \cite{matsu2}, where it is speculated that this anisotropy
does not lead to an SG phase at finite temperatures.

Another question, which is particularly significant, is the study of 
quantum effects in the theory of spin glasses \cite{sachdev}. 
From the theoretical point of view, it is well known that
quantum spin glasses, in comparison with their classical counterparts, are far
from being trivial, due to the noncommutativity of the spin operators
involved (see, for example, the discussion for the quantum
transverse Ising SG model with short- and long-range
interactions, in Refs. \onlinecite{bia} and \onlinecite{kopec1}, 
respectively). In the limit of very
low temperatures the role of quantum fluctuations in pure or
disordered systems becomes more and more important. At the critical point
itself, fluctuations exist over all scales. At moderate temperatures,
quantum fluctuations are usually suppressed in comparison with thermal ones.
At low temperatures, however, quantum fluctuations, specially in low-lying
states, may dominate and strongly influence the critical behavior of the
system. There are a
few works on quantum Heisenberg SG, but only infinite-range-interaction 
models have been treated \cite{kopec2}. 

Our motivation for this work is the well known fact that the anisotropy may
change the nature of phase transitions in a fundamental way, 
and may induce the appearance of an SG phase in the 
three-dimensional short-range Heisenberg model. Also, 
the quantum influence in the phase diagram is a matter of
intrinsic interest, particularly from the experimental point of view, with
relation to high-temperature superconductor materials \cite{bed}.
Aharony \textit{et al.} \cite{aha} suggested a mechanism in which doping 
by holes
introduces ferromagnetic bonds into an otherwise antiferromagnetic quantum
spin-$1/2$ Heisenberg model. These holes are localized in the insulating
antiferromagnetic phase and their effect can be well approximated by a
quenched random distribution of ferromagnetic bonds which display an SG phase
at low temperatures, as a window between the insulating antiferromagnetic 
phase and the
superconducting phase. Physically, the SG phase in this new superconductor
compounds is attributed to the presence of the random Dzyaloshinski-Moriya
interaction \cite{aha}. We will show in this work that the presence
of uniaxial anisotropy induces an SG phase at low
temperatures only for a finite value of the anisotropy; for small enough 
values of the anisotropy, no long-range SG order is observed,
which confirm the results of Mastubara \textit{et al.} 
\cite{matsu2}.

\section{Method}

	The main issue we want to address is the influence of a 
uniaxial anisotropy
on the phase diagram of the anisotropic quantum Heisenberg spin-glass, with
hamiltonian: 
\[ {\cal H} = - \sum_{<i,j>} J_{ij} \left[ (1-\Delta_{ij}) 
(\sigma_i^x \sigma_j^x 
+ \sigma_i^y \sigma_j^y) + \sigma_i^z \sigma_j^z \right], \]
where $\sigma_i^{\alpha}$ is the component $\alpha$ of a spin-1/2 Pauli
matrix in site $i$ and the sum is over all 
first-neighbor bonds on a cubic lattice. In this work, we study two
different initial probability distributions for $J_{ij}$, a Gaussian
and a uniform one, respectively:
\[  {\cal P}(J_{ij}) = \frac{1}{\sqrt{2 \pi \bar{J}}} \exp(-J_{ij}^2
   / 2 \bar{J}^2),   \]
   or 
  \[  {\cal P}(J_{ij}) = \left\{
  \begin{array}{cl} \frac{1}{2 \sqrt{3} \bar{J}}, &
        -\sqrt{3} \bar{J} < J_{ij} < \sqrt{3}\bar{J} \\
	0, & \mbox{otherwise},
  \end{array}  \right.
   \] 
where $\bar{J}$ is the width of the distributions. On the other hand,
the probability distribution for $\Delta_{ij}$ is, {\bf initially},
given by:
\[ {\cal P} (\Delta_{ij}) = \delta(\Delta_{ij}-\Delta_0). \]
We use a
real-space renormalization-group approach; this method has 
been successfully applied in the study of both classical and
quantum models.  The formalism is specially suitable to obtain
multidimensional
phase diagrams and qualitative results, indicating universality
classes and possible crossover phenomena. A great variety of RG methods
has been proposed \cite{mariz,qrg} over the last years and applied
with success in many different quantum systems \cite{aplic,nsb}.
Recently, an important simplification of the successful method
introduced in Ref. \onlinecite{mariz} has been proposed \cite{newrg};
we will develop even further this new approach in this work.

	Within the context of a small-cell approximation,
the simple cubic lattice is represented by
a hierarchical one \cite{hl}, depicted in Fig.~\ref{cell}.   
The use of this particular hierarchical lattice
is equivalent to a Migdal-Kadanoff approximation \cite{mk}. 
The original lattice is shown on the left-hand side of Fig.~\ref{cell},
with different interactions
$K_{ij} \equiv J_{ij}/k_B T$ and anisotropy parameters $\Delta_{ij}$ between 
first-neighbor spins $\sigma_i$ and $\sigma_j$. Performing a partial trace 
over spins $\sigma_3$, $\sigma_4$, $\sigma_5$, and $\sigma_6$, 
we obtain a renormalized
Hamiltonian, with parameters $K'_{ij}$ and $\Delta'_{ij}$
(right-hand side of Fig.~\ref{cell}).

	First, we have to calculate the renormalized distributions
(forcing back the distributions to their original shapes leads to
wrong results \cite{spinglass}). To do so,
we choose the eight interaction parameters for the
original lattice, $K_{ij}$, from the original
distribution, while all $\Delta_{ij}$ are the same, {\bf initially};
then we calculate the renormalized $K'$ and $\Delta'$.
This is done a number of times (usually of the order of 1 million),
to get new probability distributions for $K'$ and
$\Delta'$. The anisotropy parameter, although uniform in the
first iteration of the renormalization group, follows a disordered
probability distribution, afterwards. Also, the distribution for $K'$ is
no longer the same as the initial one.
For the second iteration, we choose
$K'_{ij}$ and $\Delta'_{ij}$ from the renormalized distributions
obtained in the first iteration, combine them as in the
left-hand side of Fig. 1,  and then calculate
$K''$ and $\Delta''$, i.e, the renormalized quantities for the
second iteration. This process is repeated until
we reach a "fixed-point" distribution. Alternatively, we can
choose to follow the distribution functions for $K$ and
$K^{xy} \equiv K(1-\Delta)$; we will compare below the results for
both procedures.

	For each set of $K_{ij}$ and $\Delta_{ij}$, the renormalized
quantities are calculated as follows.
Given a set of parameters, chosen from a given probability
distribution,  we impose that:
\be
 \left< m_1 m_2 | \rho' |  m_1 m_2 \right> = \nonumber \\
Tr' \left< \{m\} | \rho(\{K,\Delta\}) 
| \{m\} \right>,  \label{eq:RGE} 
\ee
where $\left. |\{m\}\right>$ stands for 
$\left. | m_1 m_2 m_3 m_4 m_5 m_6 \right>$ (and in a similar way
 for the ``bra''), $\rho(\rho')$ is the density matrix of the original
(renormalized) cell ($\rho'$ is a function of the renormalized parameters
$K'$, $\Delta'$, and $C'$), $m_i$ is the eigenvalue of the
$\sigma^z$ operator at site $i$, $Tr'$ means a partial trace over spins
$\sigma_3$, $\sigma_4$, $\sigma_5$, and $\sigma_6$, and $\{K, \Delta\}$ stands
for all sixteen parameters in the original cell. Only three elements
of $\rho'$ are non-zero, and this is the number of
renormalized quantities: $K'_{ij}$, $\Delta'_{ij}$ and $C'$ ($C'$ is a 
constant generated
by the renormalization procedure which is not relevant for obtaining
the phase diagram). So, no extra equation is needed
and the procedure is exact at the cluster level.
One great advantage of this approach is that no expansion
of the Hamiltonian is necessary; this expansion becomes cumbersome
if cells with more sites are employed or if models with
spin 1 or greater are treated. Moreover, our procedure recovers the
same recursion relations as former treatments \cite{mariz}.

\section{Results and Discussion}
	
	Our goal is to obtain the $kT/\bar{J} \times \Delta_0$ phase-diagram.
We start from many different points in this diagram and follow the 
renormalized distributions until a given attractor is reached. 
In Fig.~\ref{pd} this phase-diagram is depicted: $SG$ stands for
the spin-glass phase while $P$ stands for the paramagnetic one.
We expect the spin-glass phase, which is certainly present for 
the Ising model ($\Delta_0=1$) \cite{spinglass}, to extend for smaller 
values of $\Delta_0$. This is the case but notice that the transition line 
goes to zero at a value of $\Delta_0$ greater than zero. This behavior is 
analogous to the one for the antiferromagnetic anisotropic Heisenberg model
on the square lattice \cite{nsb}, except that in the latter model a
reentrant behavior is obtained. The fact that the transition line
does not extend to $\Delta_0=0$ is usually due to quantum fluctuations
which, at low temperatures, are important and, together with thermal
fluctuations, tend to drive the system to a disordered phase.

For the Ising sub-space ($\Delta_0=1$) the fixed-point
distribution for the paramagnetic phase attractor is such that 
$\bar{J}/kT=0$, while for the SG attractor, $\bar{J}/kT=\infty$.
There are still possible fixed points at the line $\Delta_0=0$ (isotropic
Heisenberg spin glass) but they  were not found in this work (see below).
For any $\Delta_0 \neq 0$, the attractor is found to be at the line 
$\Delta_0=1$, that is, any initial point with $\Delta_0 \neq 0$ flows, upon
application of the renormalization-group procedure, to the $\Delta_0=1$ 
subspace. Exactly at the transition line, the flow
is towards the Ising "fixed-point" (point $I$ in Fig.~\ref{pd})
and the whole line is attracted to the distribution at that point. 
Physically, this means that 
the critical behavior along the line is the same as for the Ising
spin glass.
Critical exponents for the Ising spin glass are the same as those
calculated in Ref. \onlinecite{nogueira}; moreover, the critical
probability distribution is the same as in the cited reference, for both
Gaussian and uniform probability distributions.

	Some points are worth mentioning here. The distributions for $K_{ij}$
and $\Delta_{ij}$, 
after the first iteration, do not retain its original form.
Therefore, a more complete
picture of this problem would involve a flux on a space of probability
distributions. The phase diagram we chose to represent our results is
only a schematic one. On the other hand, if $\Delta_0$ 
is different from 0 and 1, its distribution after the first iterations
is not uniform anymore. It evolves along the renormalization-group
procedure and only when the number of iterations
increases, the distribution for $\Delta$ is again a delta function,
${\cal P}(\Delta_{ij}) = \delta(\Delta_{ij}-\Delta_0)$, with $\Delta_0=1$
and zero width. 

	The phase-diagram (Fig. 2) shows that there is no spin-glass phase 
for the isotropic spin-1/2 Heisenberg model in three dimensions. This result
confirms those found in earlier works \cite{matsu1,banavar},
indicating that the lower critical dimension for the isotropic spin-1/2
Heisenberg spin glass is greater than three. On the other hand, Lee
and Young \cite{leeyoung} found a spin-glass phase for the
{\it classical} 3-D isotropic Heisenberg model. As the transition takes
place at low temperature, it is possible that quantum fluctuations,
present for the spin-1/2 model, are strong enough to eliminate the
spin-glass phase. We also find that an 
infinitesimal uniaxial anisotropy is not able to create an SG phase
in a Heisenberg spin-1/2 system, in three dimensions; this is consistent
with the findings of Ref. \onlinecite{matsu2}. 

	Our results are qualitatively the same for both Gaussian 
and uniform distributions. We have also used a correlated distribution
for $K_{ij}$ and $\Delta_{ij}$ \cite{berker} and the results suffer only
minor changes, maintaining the overall behavior. In another approach we 
followed, the probability distributions for the interactions $K_{ij}$
and $K^{xy}_{ij}$ were followed; again, the qualitative behavior is the 
same as when we follow the distributions for $K_{ij}$ and 
$\Delta_{ij}$.

	Finally, let us mention that, contrarily to what happens for 
systems where only ferromagnetic (or antiferromagnetic) interactions
are present \cite{mariz}, there is a strong difference between
treating the cell ``as a whole'' or ``by pieces''. In the latter,
the original cell (see Fig.~\ref{cell}) is seen as a combination in 
parallel of 4 interactions, each one made of two interactions in
series. In this way, the renormalized interaction and anisotropy
can be first calculate for each combination in series and then
combined in parallel. For systems with no frustration \cite{mariz},
this is shown to introduce an error smaller than 10\%, when
compared to treating the eight bonds and six spins of the cell
``as a whole''. This is no longer the case for the model we
study here and the errors are much
bigger. We believe that, for any system in which frustration
is present, the RG procedure has to done using the whole
cell. This is due to the fact that frustration is not taken into
account when the cell is renormalized by pieces.

\section{Summary}

	We applied a quantum renormalization-group procedure to the
anisotropic three-dimensional spin-1/2 Heisenberg spin glass. A 
Migdal-Kadanoff approximation is used and the $kT/\bar{J} \times \Delta_0$
phase-diagram is calculated. The spin-glass phase, 
present for the Ising model ($\Delta_0=1$) extends to smaller values of
the anisotropy parameter. The transition temperature, which separates
the ferromagnetic and paramagnetic phases, goes to zero at approximately 
$\Delta_0 = 0.59$. Acording to the approximation we used, the isotropic 
quantum Heisenberg spin glass has no spin glass phase at finite temperature.
The whole transition line between the SG phase and the paramagnetic
one is found to belong to the same universality class of the three-dimensional
Ising spin glass. Our conclusions hold true for Gaussian and uniform
distributions, for correlated distributions, and when the probability
distributions for $(K_{ij},\Delta_{ij})$ or $(K_{ij},K^{xy}_{ij})$ are
renormalized.

\begin{acknowledgments}
The authors would like to thank CNPq, CAPES, FAPERJ, FUNCITEC, and FAPEAM
for partial financial support and E. Curado and S. Coutinho for
helpful discussions.
\end{acknowledgments}

\newpage
\begin{figure} %Figure1%
\caption{Hierarchical lattice suitable for calculating
the renormalization-group transformations on the simple
cubic lattice. The calculation of the
renormalized quantities (right-hand side of figure) is explained
in the text.}
\label{cell}
\end{figure}

\begin{figure} %Figure1%
\caption{Approximate phase diagram for the anisotropic Heisenberg spin-glass
on the cubic lattice. $SG$ stands for the spin-glass phase, $P$ stands
for the paramagnetic phase, and $I$ stands for the Ising transition point.
The continuous line is a guide to the eye.}
\label{pd}
\end{figure}

\end{document}